# AgentFacts: Universal KYA Standard for Verified AI Agent Metadata & Deployment

**Jared James Grogan, Universitas AI**
jared.grogan@post.harvard.edu

## Abstract

Enterprise AI deployment faces critical "Know Your Agent" (KYA) challenges where organizations must verify third-party agent capabilities and establish trust without standardized metadata or verification infrastructure. Current approaches rely on self-declared capabilities and custom integration processes that create trust gaps and coordination friction limiting confident enterprise adoption. This paper presents AgentFacts, a universal metadata standard that enables systematic agent verification through cryptographically-signed capability declarations, multi-authority validation, and dynamic permission management. The specification introduces domain-specialized verification where different trusted authorities validate specific metadata aspects based on their expertise, eliminating single points of trust failure while enabling graduated confidence assessment. AgentFacts transforms agent procurement from custom integration projects into standardized workforce management, providing the transparency and governance infrastructure necessary for enterprise AI coordination at scale.

## 1. Introduction

AI agents require standardized metadata disclosure to enable discovery, trust assessment, and coordination across organizational boundaries. Similar to nutrition facts labels that provide consumers with standardized information about food products before purchase, agents need uniform metadata formats that allow stakeholders to evaluate capabilities, compliance status, and operational parameters before engagement. Current approaches to agent metadata fall into two inadequate categories: centralized registries that create single points of failure and control bottlenecks, or self-declared capabilities that lack independent verification and create trust gaps in enterprise environments.

The Know Your Agent (KYA) paradigm addresses the fundamental trust gap in AI agent deployment, analogous to how Know Your Customer (KYC) solved coordination challenges in financial services. AgentFacts provides the first comprehensive KYA solution through cryptographically-verified metadata that transforms agent capabilities from marketing claims into independently validated facts, enabling systematic procurement, governance, and coordination across organizational boundaries.

The fundamental challenge lies in establishing what constitutes reliable agent metadata. Agent facts are defined as verified metadata—information that has been independently validated by trusted authorities, similar to notarized affidavits or certified documents. This distinguishes agent facts from self-declared capabilities, which may be inaccurate or misleading. This represents a fundamental epistemological shift from self-claimed assertions to cryptographically-verified facts, establishing a new standard of evidence for agent capabilities. Verified metadata enables stakeholders to make informed decisions about agent deployment while reducing the verification burden on individual organizations.

AgentFacts addresses this challenge through a universal metadata standard that supports three primary stakeholder categories: enterprises requiring governance and compliance automation, consumers needing capability transparency and safety information, and governments seeking regulatory oversight and coordination mechanisms. The standard employs a multi-authority verification model where different trusted entities can validate specific aspects of agent metadata, eliminating dependence on single verification sources while maintaining cryptographic integrity.

The AgentFacts specification is released under the Apache 2.0 license as foundational internet infrastructure, similar to protocols like TCP/IP that enable distributed coordination without central control. This open standard approach facilitates global adoption while allowing commercial services and tools to build upon the specification. The standard enables peer-to-peer agent discovery without requiring centralized registries, reducing infrastructure dependencies and eliminating single points of failure that could disrupt agent coordination networks.

Dynamic metadata represents a core design principle, recognizing that agent capabilities, permissions, and compliance status change over time. Unlike static profile systems, AgentFacts incorporates time-to-live mechanisms and update protocols that ensure metadata freshness while maintaining verification integrity. This dynamic approach proves particularly valuable for enterprise agent integration scenarios where permissions and roles must adapt to changing operational requirements.

The nutrition facts metaphor extends beyond consumer applications to enterprise and government contexts. Just as food manufacturers must disclose ingredients and nutritional information to enable informed purchasing decisions, agent providers must disclose capabilities, limitations, and compliance status to enable informed deployment decisions. This transparency model reduces friction in agent adoption by providing standardized evaluation criteria that stakeholders can apply consistently across different agent providers and use cases.

## 2. Related Work

Agent coordination and discovery face challenges similar to those addressed by established metadata standards across other domains. Existing standards like OpenAPI focus on interface specification but lack the governance, compliance, and trust metadata necessary for enterprise agent adoption. Current AI agent coordination approaches fall into three categories: proprietary vendor registries that create lock-in and single points of failure, academic frameworks that lack practical implementation paths, and emerging standards like Google's A2A specification [1] that focus on agent interoperability, and ad-hoc integration approaches that require custom development for each agent relationship.

## 3. AgentFacts Specification Architecture

The AgentFacts specification defines a universal metadata schema that accommodates enterprise, consumer, and government stakeholder requirements while maintaining structural consistency. Different stakeholders prioritize different metadata aspects: enterprises focus on governance and compliance automation, consumers require capability transparency and safety information, and governments need regulatory oversight and coordination mechanisms. Rather than creating separate schemas for each context, AgentFacts employs a unified template with stakeholder-specific metadata sections that can be selectively populated and disclosed based on deployment requirements.

AgentFacts functions as a universal superset schema where enterprises might utilize all ten sections for comprehensive governance, consumers might focus primarily on baseline model and safety classifications, and governments might emphasize compliance and verification metadata. Designed from first principles as a comprehensive superset, AgentFacts can accommodate existing standards like Google's A2A Agent Cards within its capabilities framework while adding the verification, compliance, and governance layers necessary for enterprise adoption. This superset approach enables a single global standard that accommodates diverse stakeholder needs through selective implementation rather than requiring multiple competing schemas, positioning AgentFacts as the foundational metadata infrastructure for all agent coordination scenarios worldwide.

The universal schema structure consists of ten core sections encompassing identity, model transparency, classification, capabilities, authentication and dynamic permissions, compliance, performance, supply chain, verification, and extensibility mechanisms. Core identity provides unique agent identification using decentralized identifiers (DIDs) along with human-readable names and version information. Classification metadata categorizes agents by operational level, deployment scope, and stakeholder context, enabling appropriate discovery and coordination mechanisms. Capabilities metadata specifies supported interfaces, external APIs, tool calling specifications including Model Context Protocol (MCP), and domain-specific competencies. Authentication and dynamic permissions define current access levels, scope boundaries, and time-limited authorizations that can be updated in real-time. Compliance metadata maps to regulatory frameworks including EU AI Act requirements, NIST AI Risk Management Framework alignment, and sector-specific standards. Performance metadata provides measurable characteristics including latency, throughput, availability commitments, and cost structures. Verification metadata contains cryptographic signatures from multiple authorities and trust policy specifications.

Dynamic metadata with time-to-live mechanisms addresses the reality that agent characteristics change over time. Traditional static profile systems become outdated quickly and create stale information problems that undermine trust and coordination effectiveness. AgentFacts incorporates TTL values at both the overall metadata level and for individual fact categories, allowing different update frequencies for different types of information. Performance metrics may require frequent updates to reflect current operational status, while compliance certifications may have longer validity periods. The TTL mechanism includes webhook endpoints for push-based updates and cryptographic signature chains that maintain verification integrity across metadata updates.

Multi-authority cryptographic verification enables distributed trust without requiring consensus among all verification sources. Different authorities can validate different aspects of agent metadata based on their expertise and trustworthiness in specific domains. A cybersecurity firm might verify security-related capabilities while a compliance consultancy validates regulatory adherence. This distributed verification model prevents single points of trust failure while allowing stakeholders to weight different authorities based on their own trust policies. Verification signatures include authority identification, signature timestamps, verification scope, and confidence levels that help consuming systems make appropriate trust decisions.

Identity, role, and constitution assignment through verified metadata enables systematic agent onboarding across organizational boundaries. When enterprises integrate third-party agents, they receive baseline identity metadata that has been verified by trusted authorities. The organization then assigns roles and constitutions—operational parameters and behavioral guidelines—through additional metadata layers that define the agent's specific function within that organizational context. This separation between verified identity and assigned role allows the same agent to operate in different capacities across multiple organizations while maintaining consistent baseline verification.

The schema employs extensible design principles that accommodate future requirements without breaking backward compatibility. Custom fact categories can be added through extension mechanisms that preserve core schema validation while allowing domain-specific metadata. Integration hooks provide standardized interfaces for external systems including enterprise resource planning, customer relationship management, and compliance monitoring tools. Schema versioning enables evolution while maintaining interoperability with existing implementations.

Future-proof design considerations include support for emerging verification methods, accommodation of new regulatory frameworks, and integration with evolving identity and access management standards. The specification defines extension points for additional cryptographic signature schemes, alternative verification authorities, and new compliance frameworks that may emerge as AI governance evolves. This extensibility ensures that early adopters of the AgentFacts standard can continue using their implementations as requirements and technologies advance.

## 4. Minimum Viable Universal Implementation

### 4.1 AgentFacts Schema

The AgentFacts schema consists of ten extensible parent categories designed for incremental adoption across stakeholders.

> **Core Identity**: Unique identification using decentralized identifiers (DIDs), human-readable names, creation timestamps, and global TTL management.
>
> **Baseline Model**: Foundation AI model transparency including provider, version, training data sources, fine-tuning specifications, bias assessments, and safety evaluations.
>
> **Classification**: Universal categorization by agent type (assistant/autonomous/tool/workflow), operational level (ambient/supervised/autonomous), and stakeholder context (enterprise/consumer/government).
>
> **Capabilities**: Extensible capability declarations including external APIs, tool calling protocols (MCP/function calls), programming languages, data formats, and domain expertise.
>
> **Authentication & Dynamic Permissions**: Time-limited, scope-specific access control with permission authorities, escalation policies, and cryptographic audit trails.
>
> **Compliance & Regulatory:** Multi-jurisdictional regulatory compliance markers for EU AI Act, NIST AI RMF, GDPR, sector standards, and safety classifications.
>
> **Performance & Reputation:** Measurable quality metrics including latency percentiles, availability SLAs, throughput limits, accuracy scores, and user satisfaction ratings.
>
> **Supply Chain**: SBOM integration providing transparency into component dependencies, data sources, infrastructure providers, security scanning, and license compliance.
>
> **Verification:** Multi-authority cryptographic signatures with configurable trust policies, confidence levels, and revocation status management.
>
> **Extensibility:** Standardized extension mechanisms for custom facts, integration hooks, schema evolution, and backward compatibility support.

**Figure 1. -** The AgentFacts Ten Parent Categories.

The complete technical specification is provided in Appendix A. This comprehensive schema enables practical deployment through a simplified implementation approach that prioritizes adoption ease while maintaining full functionality.

## 4.2 Minimum Viable Universal Implementation

The minimum viable universal implementation prioritizes adoption simplicity while providing comprehensive metadata structure for agent coordination. The schema represents a light technical ask—simply adding standardized metadata labels to existing agents—that delivers significant benefits through universal adoption and network effects. Organizations can begin with internal agent coordination and gradually expand to cross-organizational scenarios as their requirements and capabilities evolve.

Performance metadata encompasses intrinsic, measurable characteristics including latency, throughput, availability, and accuracy metrics that can be objectively verified through technical testing. Reputation metadata captures extrinsic assessments from stakeholders including user satisfaction ratings, peer evaluations, and community feedback that reflect subjective quality judgments based on operational experience.

The flexible identity architecture accommodates different organizational maturity levels and technical infrastructure. Organizations can begin with simple UUID-based identifiers for internal agent coordination, such as managing agents within enterprise intranets or departmental AI deployments. As coordination requirements expand beyond organizational boundaries, they can migrate to URI-based identifiers that leverage existing domain authority, or adopt decentralized identifiers for full peer-to-peer coordination capabilities. This progression path enables incremental adoption without requiring initial investment in complex identity infrastructure.

Internal adoption scenarios demonstrate the universal value proposition through immediate organizational benefits. An enterprise can deploy AgentFacts metadata for internal agents to improve governance, enable better resource allocation, and streamline compliance reporting. These internal coordination benefits create immediate return on investment while building organizational capabilities for eventual cross-organizational coordination. The nutrition facts approach proves equally valuable for internal transparency, helping different departments understand agent capabilities and limitations before requesting agent services.

The baseline model section addresses critical transparency requirements that benefit both internal and external coordination scenarios. For internal deployments, knowing the foundational AI model helps organizations track technology dependencies, assess security implications, and plan for model updates or migrations. For external coordination, baseline model transparency enables informed decision-making about agent capabilities, potential biases, and regulatory compliance requirements. This transparency proves particularly valuable as model-specific regulations emerge and organizations need to demonstrate compliance with AI governance requirements.

Dynamic permissions architecture provides immediate value for internal agent management by solving the over-privileging problems common in enterprise environments. Organizations can implement time-limited, scope-specific permissions for internal agents without requiring complex external verification infrastructure. As coordination expands beyond organizational boundaries, the same permission metadata structure supports multi-party governance scenarios where different organizations need to coordinate agent access and capabilities.

Compliance and regulatory metadata provides significant value through automation of documentation and reporting requirements. Organizations benefit from standardized compliance tracking even for internal agent deployments, as regulatory requirements increasingly apply to AI systems regardless of their deployment context. The multi-jurisdictional design enables organizations to prepare for regulatory expansion while maintaining current compliance with existing requirements.

The universal standard approach could deliver network effects and marginal benefits that increase with adoption scale. Early adopters benefit from internal coordination improvements and regulatory automation. As adoption expands, organizations gain access to broader agent ecosystems, reduced integration complexity, and enhanced coordination capabilities. The light implementation requirement—simply adding metadata labels to existing agents—ensures that adoption friction remains minimal while benefits compound through network participation.

SBOM integration provides supply chain transparency that proves valuable for both internal governance and external coordination. Organizations benefit from understanding their agent dependencies for security, compliance, and risk management purposes. This transparency becomes increasingly important as AI systems become more complex and regulatory requirements expand to cover AI supply chain management.

## 5. Employee Agent Application Example

Employee agents represent AI systems that organizations integrate as workforce members with ongoing roles, responsibilities, and governance oversight, distinct from temporary tool usage or one-time task execution. Unlike traditional software services, employee agents require identity establishment, role assignment, permission management, and performance evaluation similar to human workforce management processes

The employee agent paradigm represents a significant shift in enterprise AI adoption, where organizations integrate third-party agents as workforce members rather than temporary tools. This section demonstrates how AgentFacts enables systematic onboarding, role assignment, and governance of agent employees through a concrete enterprise scenario.

Consider a financial services organization that contracts with a specialized AI provider to deploy a financial analysis agent for regulatory reporting tasks. The agent provider supplies an AgentFacts metadata package that includes verified baseline model information indicating the agent runs on a fine-tuned version of a leading foundation model with specialized training on financial regulations and reporting standards. The verification section contains cryptographic signatures from the AI provider, a financial compliance consultancy that validated regulatory knowledge, and a cybersecurity firm that assessed security controls.

The enterprise onboarding process begins with AgentFacts metadata evaluation. The compliance section indicates EU AI Act classification as a limited-risk system with appropriate transparency obligations, NIST AI Risk Management Framework alignment for financial applications, and certifications for relevant standards including SOX compliance and ISO 27001 security controls. The capabilities metadata specifies support for financial data formats, integration with common enterprise resource planning systems, and tool calling capabilities that include Model Context Protocol access to approved financial databases and reporting tools.

Role assignment occurs through metadata layering where the enterprise adds organization-specific metadata to the verified baseline facts. The classification section receives updates to specify stakeholder context as enterprise and deployment scope as internal, while dynamic permissions metadata defines the agent's specific operational boundaries within the organization. Initial permissions include read access to designated financial databases, write access to draft reporting templates, and execute permissions for approved analytical tools. The scope of work metadata specifies boundaries around quarterly regulatory reporting tasks with explicit exclusions for real-time trading decisions or customer data access.

Dynamic permission management addresses the over-privileging problem through time-limited, scope-specific access control. During the initial quarter, the agent receives broad read permissions for historical financial data to establish baseline analytical capabilities. As the quarterly reporting deadline approaches, permissions automatically expand to include write access to official reporting systems, but only for specific document templates and with mandatory human review requirements. After report submission, permissions automatically revert to the baseline level, preventing unnecessary access accumulation that creates security risks.

The permissions architecture incorporates temporal and contextual constraints that reflect real-world operational requirements. For example, the agent receives elevated permissions during business hours when human supervisors are available for escalation, but operates with restricted permissions during off-hours when oversight is limited. Geographic restrictions ensure the agent only accesses data from approved jurisdictions, supporting compliance with data residency requirements. Permission changes generate audit trail entries that document who authorized modifications, when they occurred, and what business justification supported the change.

Governance capabilities leverage AgentFacts metadata to provide comprehensive oversight and control mechanisms. The performance and reputation section receives regular updates documenting the agent's accuracy in generating regulatory reports, response times for analytical queries, and user satisfaction ratings from human colleagues. This performance data enables evidence-based decisions about contract renewal, scope expansion, or remediation requirements. Compliance metadata automatically updates as new regulatory requirements emerge, ensuring the agent maintains appropriate certifications and controls.

Audit capabilities benefit from the structured metadata format and cryptographic verification mechanisms. Internal audit teams can review permission histories, performance trends, and compliance status through standardized interfaces rather than requiring custom integrations with each agent provider's proprietary systems. External auditors can verify agent controls and capabilities through the cryptographically signed metadata, reducing the audit burden while increasing confidence in agent-generated outputs.

The supply chain transparency provided through SBOM integration proves particularly valuable in regulated industries where third-party risk management requires comprehensive vendor assessment. The financial services organization can evaluate the agent's dependency on cloud infrastructure providers, assess the security scanning results for underlying software components, and verify that all supply chain elements meet appropriate security and compliance standards.

Multi-jurisdictional compliance automation demonstrates particular value when the financial services organization expands operations to new regions. The agent's metadata includes compliance markers for multiple jurisdictions, enabling deployment in EU, US, and Asia-Pacific markets without requiring separate agent procurement and onboarding processes. As regulatory requirements change, compliance metadata updates automatically trigger review processes and permission adjustments to maintain ongoing compliance.

This employee agent scenario illustrates how AgentFacts transforms third-party AI integration from a complex, custom integration challenge into a standardized workforce management process. The verified metadata provides the transparency and control mechanisms that enterprises require for confident agent adoption while reducing the integration complexity and ongoing management overhead associated with agent workforce coordination.

## 6. Technical Implementation Considerations

Cryptographic signatures ensure metadata integrity and authenticity across distributed verification scenarios. AgentFacts employs a multi-signature architecture where different verification authorities can independently sign specific metadata sections using standard cryptographic algorithms including ECDSA, RSA, and emerging post-quantum resistant schemes. Each signature includes the signing authority's identifier, timestamp, verification scope indicating which metadata sections were validated, and confidence level expressing the authority's assessment certainty. Signature verification requires access to authority public keys and certificate chains, supporting both centralized certificate authorities and decentralized web-of-trust models depending on deployment requirements.

The verification process accommodates partial trust scenarios where consuming systems may trust different authorities for different types of metadata validation. A healthcare organization might weight medical compliance certifications more heavily than general security assessments, while a financial services firm might prioritize regulatory compliance over performance metrics. The technical architecture supports configurable trust policies that allow organizations to specify minimum signature requirements, acceptable verification authorities, and confidence thresholds for different operational contexts.

TEE and confidential computing integration addresses privacy requirements in multi-party coordination scenarios where agents must interact across organizational boundaries while protecting sensitive information. The AgentFacts specification includes metadata fields for TEE capabilities, supported confidential computing platforms, and privacy-preserving protocols that enable selective disclosure. Agents operating within trusted execution environments can provide verified attestations of their privacy controls while maintaining confidentiality of proprietary algorithms, training data, or operational parameters. This capability proves particularly valuable for enterprise scenarios where competitive advantages must be protected while enabling necessary coordination and verification.

Confidential computing integration supports secure multi-party computation scenarios where multiple agents collaborate on shared tasks without revealing proprietary information to each other or to external observers. The metadata specification includes references to supported protocols, key management approaches, and attestation mechanisms that enable consuming systems to verify privacy protections before engaging in sensitive operations. TEE attestation metadata provides cryptographic proof that agents operate within appropriate hardware security boundaries and follow verified privacy-preserving protocols.

Update mechanisms and TTL management balance metadata freshness with verification integrity through distributed update protocols that maintain cryptographic authenticity. The technical architecture supports both push-based updates through webhook notifications and pull-based refresh cycles based on TTL expiration. Update notifications include cryptographic proofs that link new metadata to previous versions, ensuring continuity and preventing unauthorized modifications. Different metadata sections can have different TTL values reflecting their update frequency requirements: performance metrics might refresh every few minutes while compliance certifications could remain valid for months.

TTL management includes graceful degradation mechanisms that handle partial metadata expiration without completely invalidating agent facts. Critical metadata sections like identity and compliance information might have longer TTL values with strict verification requirements, while operational metadata like current permissions or performance metrics can refresh more frequently with lower verification overhead. The architecture includes configurable staleness policies that allow consuming systems to specify acceptable metadata age for different operational contexts.

Interoperability with existing corporate schemas requires translation mechanisms and API compatibility layers that enable AgentFacts adoption without requiring complete replacement of existing agent management systems. The specification defines standard mapping procedures for common enterprise schemas including those used by major cloud providers, identity management systems, and compliance frameworks. JSON-LD formatting provides semantic compatibility with existing metadata systems while maintaining structural consistency across different implementation environments.

Corporate schema integration includes bidirectional translation capabilities where existing agent metadata can be automatically converted to AgentFacts format and AgentFacts metadata can be consumed by legacy systems through standard APIs. This interoperability reduces adoption friction by allowing organizations to implement AgentFacts incrementally without disrupting existing operational processes. The specification includes extension mechanisms that accommodate proprietary metadata fields while maintaining core compatibility requirements.

Performance and scaling characteristics address the distributed coordination requirements inherent in multi-agent environments. AgentFacts metadata packages are designed for efficient caching, compression, and distribution across network boundaries. Standard compression algorithms can achieve significant size reductions for typical metadata packages, while caching strategies enable local storage of frequently accessed agent facts to reduce network overhead. The specification includes performance guidelines for metadata size limits, signature verification overhead, and update frequency constraints that support large-scale deployments.

Scaling considerations include network partition tolerance and eventual consistency models that enable continued operation when some verification authorities or network segments become temporarily unavailable. The metadata architecture supports offline verification scenarios where previously cached signatures remain valid even when real-time authority verification is impossible. This resilience proves essential for enterprise environments where network connectivity cannot be guaranteed and operational continuity requires robust degradation mechanisms.

Implementation complexity management focuses on providing clear integration pathways for different organizational maturity levels. Organizations can begin with basic schema validation and progress to full multi-authority verification as their requirements and capabilities evolve. The specification includes reference implementations for common programming languages and integration patterns that reduce development overhead while maintaining security and interoperability requirements.

## 7. Adoption and Ecosystem Development

Apache 2.0 licensing positions AgentFacts as foundational internet infrastructure rather than proprietary technology, enabling broad adoption without vendor lock-in concerns that frequently impede enterprise standard adoption. The permissive license allows organizations to implement, modify, and distribute AgentFacts-compatible systems without royalty obligations or restrictive licensing terms. This approach follows successful precedents in internet protocol development where open standards like TCP/IP, HTTP, and JSON achieved universal adoption through unrestricted availability and community-driven evolution.

The open standard approach reduces adoption friction by eliminating licensing negotiations, compliance audits, and vendor dependency assessments that slow enterprise technology adoption. Organizations can implement AgentFacts validation and generation tools using internal development resources without requiring external licensing agreements. Commercial tool providers can build AgentFacts-compatible products and services while maintaining competitive differentiation through implementation quality, additional features, and service delivery rather than through proprietary standard control.

Verification authority ecosystem development leverages market-driven specialization where different organizations can establish credibility in specific verification domains. Cybersecurity firms can focus on security and privacy validation, compliance consultancies can specialize in regulatory adherence verification, and performance testing organizations can provide capability and reliability assessments. This distributed verification model prevents single points of trust failure while enabling specialized expertise development that improves overall verification quality and coverage.

The multi-authority model creates economic incentives for verification quality through competitive reputation mechanisms. Organizations seeking agent verification can select authorities based on their track record, expertise, and credibility in relevant domains. Poor verification quality damages authority reputation and reduces market demand for their services, while accurate and thorough verification builds authority credibility and market position. This market-driven quality control provides more robust incentives than centralized verification monopolies that lack competitive pressure for continuous improvement.

Standards-based compliance automation delivers significant value through reduced regulatory burden and accelerated agent deployment timelines. Organizations can leverage AgentFacts compliance metadata to automatically generate regulatory documentation, streamline audit processes, and demonstrate adherence to multiple regulatory frameworks through standardized formats. EU AI Act compliance becomes more manageable when agent capabilities, risk assessments, and transparency obligations are documented in standard formats that regulatory tools can automatically process and validate.

Multi-jurisdictional compliance automation proves particularly valuable for organizations operating across multiple regulatory environments. Rather than maintaining separate compliance documentation for each jurisdiction, organizations can leverage AgentFacts metadata that includes compliance markers for relevant regulatory frameworks. As new regulations emerge or existing requirements change, compliance metadata updates automatically trigger appropriate review and adjustment processes without requiring manual documentation maintenance.

Implementation pathway strategies accommodate different organizational maturity levels and technical capabilities. Organizations can begin with basic AgentFacts metadata generation and validation using simple tools and progress to full multi-authority verification as their requirements and capabilities evolve. Early adopters might focus on internal agent management scenarios before expanding to cross-organizational coordination. The incremental adoption approach reduces initial investment requirements while providing clear value at each implementation stage.

Community development through open source collaboration enables shared tool development, best practice documentation, and standard evolution based on real-world deployment experience. The Apache 2.0 license encourages community contributions including reference implementations, integration libraries, and domain-specific extensions that benefit all adopters. Community-driven development also ensures that standard evolution reflects diverse stakeholder requirements rather than single vendor priorities.

The ecosystem development model anticipates commercial service opportunities built upon the open standard foundation. Organizations can provide verification authority services, compliance automation tools, agent discovery platforms, and integration consulting while leveraging the shared AgentFacts standard for interoperability. This approach creates market opportunities for specialized service providers while maintaining open standard benefits for adopting organizations.

Global adoption potential stems from the universal applicability of standardized agent metadata across different industries, regulatory environments, and technical architectures. The nutrition facts metaphor translates across cultural and regulatory boundaries while the technical architecture accommodates different cryptographic standards, compliance frameworks, and operational requirements. This universality may enable network effects where increasing adoption improves value for all participants through enhanced interoperability and reduced integration complexity.

AgentFacts provides the foundational metadata standard necessary for confident AI agent adoption across organizational boundaries. The combination of verified metadata, multi-authority trust, and open standard accessibility addresses the transparency, security, and interoperability requirements that currently limit enterprise agent integration. Through systematic adoption and community development, AgentFacts could enable the agent coordination capabilities necessary for the next phase of AI deployment at scale.

## APPENDIX A.

**AGENTFACTS MINIMUM VIABLE UNIVERSAL SCHEMA (v1.0)**

**CORE IDENTITY (Required)**

- **agent_id:** Unique identifier (DID, UUID, URI, etc.)
- **name:** Human-readable agent identifier
- **version:** AgentFacts schema version
- **created:** ISO timestamp of initial registration
- **last_updated:** ISO timestamp of most recent update
- **ttl:** Global time-to-live in seconds

**BASELINE MODEL (Required)**

- **foundation_model:** Base AI model (gpt-4, claude-3, etc)
- **model_version:** Specific version identifier
- **model_provider:** Organization providing base model
- **training_data_sources:** Data used for fine-tuning
- **training_cutoff_date:** Knowledge cutoff timestamp
- **fine_tuning:** Custom training specifications
- **model_capabilities:** Multimodal, reasoning, tool use
- **known_limitations:** Documented capability boundaries
- **bias_assessments:** Fairness and bias evaluation results
- **safety_evaluations:** Risk and safety testing outcomes

**CLASSIFICATION (Universal Categories)**

- **agent_type:** [assistant|autonomous|tool|workflow]
- **operational_level:** [ambient|supervised|autonomous]
- **stakeholder_context:** [enterprise|consumer|government]
- **deployment_scope:** [internal|external|hybrid]
- **interaction_mode:** [synchronous|asynchronous|batch]

**CAPABILITIES (Extensible)**

- **external_apis:** List of supported API specifications
- **tool_calling:** [MCP|function_calls|custom_protocols]
- **programming_languages:** [python|javascript|sql|...]

- **data_formats:** [json|csv|pdf|image|audio|video]
- **interface_types:** [text|voice|gui|api]
- **domain_expertise:** [finance|healthcare|legal|...]
- **language_support:** [en|es|fr|de|zh|ja|...]

**AUTHENTICATION & DYNAMIC PERMISSIONS (Enterprise Critical)**

- **supported_methods**: [oauth2|api_key|mtls|jwt|saml]
- **primary_scheme:** Recommended authentication method
- **oauth_endpoints:** Authorization server configurations
- **token_requirements:** Scope and claim specifications
- **auth_security_level:** [basic|standard|high|critical]
- **session_management:** Timeout and refresh policies
- **multi_factor_required:** Boolean MFA requirement
- **auth_compliance:** [fido2|pkce|zero_trust|...]
- **current_permissions:** [read|write|execute|admin]
- **permission_scope:** Resource pattern specifications
- **scope_of_work:** Operational boundary definitions
- **permission_ttl:** Expiration timestamp for access
- **permission_authority:** Granting organization identifier
- **escalation_policy:** Permission upgrade procedures
- **audit_trail:** Permission change history

**COMPLIANCE & REGULATORY (Multi-jurisdictional)**

- **eu_ai_act:** [risk_level|transparency_obligations]
- **nist_ai_rmf:** [framework_alignment|risk_categories]
- **gdpr_compliance:** [data_protection|privacy_controls]
- **sector_standards:** [nist|iso27001|hipaa|pci_dss|...]
- **geographic_compliance:** [us|eu|uk|ca|au|sg|...]
- **safety_classification:** [low|medium|high|critical]
- **audit_certifications:** [soc2|iso|fips|...]

**PERFORMANCE & REPUTATION (Measurable Quality)**

- **response_time_p50:** Median latency in milliseconds
- **response_time_p95:** 95th percentile latency
- **availability_sla:** Uptime percentage commitment
- **throughput_limit:** Requests per minute capacity
- **accuracy_metrics:** Domain-specific performance scores
- **error_rate:** Failure percentage over time window

- **cost_structure:** [per_request|subscription|hybrid]
- **reputation_score:** Aggregate quality rating
- **user_satisfaction:** Average satisfaction ratings
- **historical_performance:** Time-series performance data

---

**SUPPLY CHAIN (SBOM Integration)**

- **component_dependencies:** Third-party service list
- **data_sources:** Training and operational data origins
- **infrastructure_providers:** Hosting and compute services
- **software_libraries:** Code dependencies and versions
- **security_scanning:** Vulnerability assessment results
- **license_compliance:** Open source license obligations
- **supply_chain_attestation:** Verification of components

---

**VERIFICATION (Cryptographic Trust)**

- **signatures:** Multi-authority cryptographic proofs
- **verification_authorities:** List of trusted validators
- **verification_policy:** Trust requirements specification
- **confidence_levels:** Authority-specific trust scores
- **verification_ttl:** Fact freshness expiration times
- **signature_algorithms:** Cryptographic methods used
- **revocation_status:** Certificate validity indicators

---

**EXTENSIBILITY (Future-Proofing)**

- **custom_facts:** Domain-specific metadata extensions
- **integration_hooks:** Webhook endpoints for updates
- **schema_extensions:** References to additional schemas
- **plugin_interfaces:** Standardized extension mechanisms
- **backward_compatibility:** Version migration support